\documentclass[12pt,letterpaper,hyper]{JHEP3}
\usepackage[utf8]{inputenc}
\usepackage{epic,eepic}
\usepackage{amsmath,epsfig}
\usepackage{amssymb,amsfonts}
\usepackage{graphicx}

\title{Stringy effects in black hole decay}

\preprint{}
\author{Stanislav Kuperstein and Sameer Murthy\\

\it {Laboratoire de Physique Th\'eorique et Hautes
Energies (LPTHE)\\
\it{Universit\'e Pierre et Marie Curie -- Paris 6; CNRS UMR
7589}\\
\it{Tour 13-14, 4$^{\grave{e}me}$ \'etage, Boite 126, 4 Place
Jussieu}, 
\it {75252 Paris Cedex 05, France}}\\

\texttt{e-mails}: \textsf{skuperst, smurthy @lpthe.jussieu.fr}\\
}

\abstract{
We compute the low energy decay rates of near-extremal three(four) charge black holes in five(four) dimensional 
$\CN=4$ string theory  to sub-leading order in the large charge approximation. This involves studying 
stringy corrections to  scattering amplitudes of a scalar field off a black hole. We adapt and use recently developed 
techniques to compute such amplitudes as near-horizon quantities.  
We then compare this with the corresponding 
calculation in the microscopic configuration carrying the same charges as the black hole. 
We find perfect agreement between the microscopic and macroscopic calculations;  
in the cases we study, the zero energy limit of the scattering cross section  is equal to four times the 
Wald entropy of the black hole.
}


\newcommand{\CA}{\mathcal{A}}

\newcommand{\CE}{\mathcal{E}}

\newcommand{\CL}{\mathcal{L}}

\newcommand{\CN}{\mathcal{N}}

\newcommand{\IR}{\mathbb{R}}

\newcommand{\half}{\frac{1}{2}}
\newcommand{\ndt}{\noindent}

\newcommand{\pa}{\partial}

\newcommand{\0}{{\scriptscriptstyle{(0)}}}
\newcommand{\1}{{\scriptscriptstyle{(1)}}}
\newcommand{\2}{{\scriptscriptstyle{(2)}}}

\def\bea{\begin{eqnarray}}
\def\eea{\end{eqnarray}}
\def\be{\begin{equation}}
\def\ee{\end{equation}}
\def\ba{\begin{align}}
\def\ea{\end{align}}
\def\bse{\begin{subequations}}
\def\ese{\end{subequations}}

\newcommand{\bem}{\begin{pmatrix}}
\newcommand{\eem}{\end{pmatrix}}

\def\={\;  = \;}
\def\+{\, + \,}

\def\wt{\widetilde}

\def\rt2{\sqrt{2}}

\DeclareMathOperator*{\Res}{Res}

\def\s{\sigma}

\def\a{\alpha}

\def\d{\delta}
\def\k{\kappa}

\def\o{{\omega}}
\def\w{{\omega}}
\def\O{{\Omega}}
\def\G{\Gamma}

\begin{document}

\section{Introduction and summary}

One of the spectacular successes of string theory has been the development of the idea that a black hole 
can be viewed as a bound state of extended objects -- strings and branes wrapped around various compact 
cycles of spacetime. At large string coupling, the Schwarschild 
radius of these objects is larger than the Compton wavelength and they are described as black hole solutions 
to the effective action of string theory. At small string coupling, 
the small fluctuations of these branes  can be described using field 
theoretic methods. 
Much of the quantitative  development of this idea has happened via the comparison of physical quantities 
computed using different methods in the two limits. In this manner, one has gained an understanding of the 
quantum properties of black holes through their description as a collection of microstates.

The most developed calculations have been those of the entropy of extremal black holes, for which one has now an 
extremely detailed understanding in supersymmetric situations as arising from a counting of BPS states carrying 
a certain set of charges, in the limit that the charges are large. Indeed, starting from the work of \cite{Strominger:1996sh} where 
the first such counting was performed with numerical agreement, the understanding has now developed much 
beyond the leading asymptotic agreement of the degeneracy. One now understands sub-leading perturbative 
corrections to the leading asymptotic degeneracy in terms of higher derivative corrections to the string effective 
action \cite{LopesCardoso:1998wt} and the application of the more general Wald formula which replaces the  Bekenstein-Hawking area law (see \cite{Sen:2007qy} and references therein). 
In situations with $\CN=4$ supersymmetry, these calculations have been pushed to non-trivial regimes where all 
but one of the charges are finite \cite{Dabholkar:2010}. 
One can even compute non-perturbative contributions to the black hole entropy \cite{Banerjee:2008ky, Murthy:2009dq}.

This spectacular agreement of the black hole entropy calculations comes about because it is essentially governed 
by an underlying supersymmetric 
index \cite{Dabholkar:2010} which does not change under continuous deformations of the moduli\footnote{There could of course be 
discrete jumps related to the appearance of multi-centered black hole bound states, how to single out the 
single-centered  black holes out of the microscopic index is an interesting question which has can be answered in 
special circumstances \cite{Dabholkar:2010mm}.}. 
For a general physical quantity, there is no {\it a priori} reason that the calculations done in the microscopic and the 
macroscopic regimes should agree.

A more dynamical observable is \emph{the rate of decay} of black holes due to Hawking radiation. 
This quantity is inherently non-BPS because in order to decay the black hole must be 
excited away from extremality, {\it i.e.} be at a non-zero temperature. 
This quantity was computed for string theoretic black holes \cite{Dhar:1996vu} and 
the microscopic and macroscopic  decay rates of near-extremal black holes were also found to agree 
at low energy to leading 
order in the large charge approximation \cite{Das:1996wn}. 
This agreement was then understood using a non-renormalization theorem on the supersymmetric 
moduli space of D-branes   \cite{Maldacena:1996iz}, and then extended to a larger 
regime of parameter space \cite{Maldacena:1997cg}. 

It is an interesting question whether this agreement of the decay rate of black holes and D-branes holds beyond 
the leading large charge approximation. In this paper, we answer this question in the affirmative for a 
certain class of black holes in $\CN=4$ string theory.

\vspace{0.2cm}

Since the area of the horizon scales as a positive power of the charges, these corrections 
can be viewed as finite size effects to the scattering problem. 
On the macroscopic side, this involves the addition of higher derivative terms to the effective 
action of string theory suppressed by the string scale $\ell_{s}$, and the computation of the black hole decay rate in this  
four derivative theory.
On the microscopic side, the problem involves a 
statistical ensemble of the states of a conformal field theory wherein the large charge limit corresponds to the 
thermodynamic limit. The corrections involve a detailed computation of the finite size effects in this statistical system. 


\subsection{The macroscopic calculation}

The quantum mechanical decay rate of black holes was first computed by Hawking \cite{Hawking:1974sw}. 
For an observer at asymptotic infinity, the rate of decay into a field quantum of a given frequency is given by 
a purely thermal factor multiplied by the {\it grey-body factor} which arises due to the effective  potential that 
the black hole creates outside the horizon for the propagation of a test particle in the background geometry 
of a black hole. This frequency dependent grey body factor is equal to the cross section of absorption $\s_{\rm abs}(\o,k_{i})$ 
of the corresponding field quantum with energy-momentum $(\o,k_{i})$ by the black hole. 
By the optical theorem, this cross section is proportional to the imaginary part of a two point function 
of the quantum field.  

The zero-frequency limit of this absorption cross section 
$\s_{\rm abs} \equiv \lim_{k_{\mu} \to 0} \s_{\rm abs}(k_{\mu})$ 
is an interesting quantity that is relevant in many physical contexts. 
For a minimally coupled scalar field in Einstein gravity, it is equal to the horizon area of the black hole, 
a result known to be universally true, independent of  the specifics of the solution 
 (similar results are known for higher spin fields) \cite{Das:1996we}.

We would now like to compute the scattering cross section of a black hole in asymptotically flat space in a higher 
derivative theory of gravity. There are many issues to deal with here, both technical and conceptual, 
some of which are discussed in \cite{Moura:2006pz}. 
We shall, however, use a different method of computation in this paper which we describe below. 

At the technical level, a notable feature of the result of  \cite{Das:1996we} is that the low energy cross section is 
equal to a quantity (the area) which can be determined by a calculation performed purely at the horizon, 
and one wonders if this is true more generally. 
For black holes in two-derivative gravity in asymptotically \emph{anti de Sitter} spaces, it has been 
shown \cite{Iqbal:2008by} that a quantity similar to the cross section can be evaluated at the horizon. 
It is also known that this feature persists in higher derivative theories of gravity 
in AdS \cite{Paulos:2009yk}, and the low frequency cross sections can always be computed 
by an effective coupling in the Lagrangian evaluated 
at the horizon \cite{Buchel:2008ae}, \cite{Brustein:2008cg}, \cite{Banerjee:2009wg}, \cite{Myers:2009ij}. 
An elegant algorithm for computing these couplings for any Lagrangian was developed in 
\cite{Paulos:2009yk}. This method which gives a result reminiscent of the Wald entropy formula 
was called the {\it pole method}. 
In this paper, we shall use this method, suitably adapted to flat space, to compute the low energy 
cross section of near-extremal black holes.

Our macroscopic analysis already throws up an interesting result. 
In the two derivative case, the cross section was equal to the horizon area 
which is equal to four times the Bekenstein-Hawking entropy. The two-derivative answer can be interpreted 
as saying that the scattered particle (or wave) interacts with each degree of freedom with equal probability, 
and the cross section simply measures the number of effective degrees of freedom, all of which live at the horizon. 
Once we go beyond the two derivative theory, we can now longer trust geometric concepts fully, and many 
geometric concepts receive stringy corrections\footnote{Indeed, this has been observed in many examples in string 
theory like in orbifolds, or the metric of moduli spaces \cite{Polchinski:1998rq}.}. A natural  guess for the scattering 
cross section that would generalize to higher theories of gravity will therefore not be the area, but  (four times) 
the entropy of the black hole, a concept which is believed to persist in any theory of gravity. 

In the examples that we consider, namely near-extremal three (four) charge black holes in five (four) dimensional 
classical superstring theory, we find that  
\emph{at very low energies the scattering cross section of a scalar field 
is equal to four times the Wald entropy 
of the black holes}. We find this result intriguing and we shall make some comments on it in the discussion at the end.

\subsection{The microscopic calculation}

The microscopic model of the black hole is that of a system of D-branes with open string excitations carrying 
momentum and energy. When the system is away from extremality, it can decay by a process in which 
open strings combine into a closed 
string and escape to infinity. At low energies, this process is dominated by a process where two open strings 
become a massless closed string mode.  In the infinite charge limit, one recovers the black hole grey body factor  
from a decay rate calculation of this D-brane system in flat space \cite{Das:1996wn}. 

This surprising result was understood as arising from the thermal nature of the open string fluctuations of the 
slightly excited D-brane system. In the low energy limit, the relevant fluctuations are the excitations of a 
$1+1$-dimensional superconformal field theory. When the momentum on the string {\it i.e.} the number of 
excitations of the SCFT is small compared to the central charge  (the {\it dilute gas approximation}), the interactions 
of the left and right movers can be neglected, and the decay rate is proportional to the entropy of the system.  
In the infinite-charge,  zero-frequency limit, the cross section is simply the extremal black hole entropy  in Planck units 
$\s_{\rm abs} = 4 G_N S$.

To include finite size effects, one crucial issue is the choice of thermodynamic ensemble. For computations of 
the quantum extremal black hole entropy, the $AdS_{2}$ boundary conditions force the system into 
the microcanonical ensemble where  all the charges and energy of the black hole are fixed \cite{Sen:2008vm}. 
For our decay rate problem, the black hole interacts with the external environment, and it is not a priori clear 
which ensemble to use. It turns out quite simply that  the $\o \to 0$ limit forces us to consider the system in the 
canonical ensemble with fixed temperature. 
Given that the gravity calculation is also naturally done in the canonical ensemble, it is then natural to make a 
comparison between the two.

We find that the cross-section of absorption of a scalar bulk mode in the $\o \to 0$ limit in the systems we 
consider is given by the same formula $\s_{\rm abs} = 4 G_N S$, but now $S$ is the full microscopic entropy 
of the system including the finite charge corrections. We note here that although we derive this formula for a 
specific system in string theory, from our derivation it is clear that the form of the answer only depends on 
there being an effective string description at low energies.

\subsection{String theory setup}

In this paper, we shall consider supersymmetric black holes in $\CN=4$ string theory compactified to four and five 
dimensional flat space. These black holes preserve $1/4$ of the supersymmetries of the theory. They have been 
studied extensively in the context of entropy counting \cite{Sen:2007qy}. The $\CN=4$ theories have a large number of 
scalar and vector fields which are  turned on in a generic $1/4$ BPS solution. One can however use the large 
duality group of the theory to obtain a simple representation of the system in a certain duality frame. 

In the four-dimensional theory, the simplest system with non-zero horizon area carries four types of charges. 
In the type IIB duality frame, the spacetime is $\IR^{3,1} \times S^{1} \times \wt S^{1} \times K3$. 
The black hole carries charges corresponding to $Q_{1}$ D1-branes wrapped on $S^{1}$, $Q_{5}$ D5-branes 
wrapped on $K3 \times S^{1}$, $n$ units of momentum on $S^{1}$ and a KK monopole corresponding to the 
$\wt S^{1}$ circle. 

In the five dimensional theory, the circle $\wt S^{1}$ becomes non-compact. The black hole now has 
three charges and is related to the 
above one by the $4d$-$5d$ lift \cite{Gaiotto:2005gf}. 
This is the  well-studied D1-D5-P system on a $K3 \times S^{1}$ compactification of IIB string theory 
 \cite{Strominger:1996sh}. 
We shall also consider these systems in the dual heterotic frame. 

The macroscopic theory is described by the effective action of string theory to four derivative level
\cite{Gross:1986mw, Metsaev:1987zx,Hull:1987pc}.
The microscopics of these systems are also understood in great detail. In particular, the degeneracy (in reality 
a supersymmetric  index)
of the $1/4$ BPS states have explicit exact formulas. In the $4d$ theory, this was first written down in \cite{Dijkgraaf:1996it}, 
and was derived more recently in \cite{Gaiotto:2005gf, David:2006yn}. The degeneracy of the $5d$ theory is closely 
related and explicit formulas have been known for some time \cite{Dijkgraaf:1996xw}.

To go from the microscopic to macroscopic pictures, one dials a single parameter which is  
the type IIB string coupling $g_{s}$ at infinity. In both 
the regimes, the string coupling remains small, but the 't Hooft couplings $g_{s}^{2}n, g_{s}^{2} Q_{1} Q_{5}$ 
are small in the microscopic D-brane picture and large in the macroscopic  supergravity picture. 

As mentioned above, the microscopic decay rate calculation in the canonical ensemble essentially reduces to a calculation 
of the entropy including the finite charge effects. In the infinite charge limit, one could use Cardy's formula to 
estimate the degeneracy of states. In our problem, we are faced the problem of estimating the degeneracy of 
states in the dilute gas region when $n \ll Q_{1} Q_{5}$ in the D1-D5 system in the 
type IIB theory  {\it i.e.} a highly off-Cardy limit. One can 
dualize the system to bring it into a Cardy limit in the heterotic frame, 
but in this case, the exact SCFT description at low energies is not known\footnote{One typically encounters 
NS5-branes in the dual system for which one does not know how to compute the density of states to the required order of 
precision.}. 
However, precisely this problem was encountered and overcome for the entropy 
\cite{Castro:2008ys, Banerjee:2008ag}, and we can safely borrow the corresponding results for the entropy estimates. 
This was done using the large group of  modular transformations underlying the system  in the $\CN=4$ situation.  
We shall always work in this regime of charges in this paper, sometimes referred to as the {\it heterotic Cardy limit}. 

\vspace{0.2cm}

To summarize, we find that in the same regime of charges $1 \ll n \ll Q_{1} Q_{5}$, both the microscopic and 
macroscopic calculations give: 
\be
\s_{\rm abs} = 4 G_N S \, , 
\ee
where $S$ is the Wald entropy of the black hole.

\subsection*{Organization of the paper}

In section \S\ref{BHreview}, we discuss the string theory setup and the corresponding black hole solutions. 
We review the four-derivative effective action of string theory. 
We review the classical solutions, their near-horizon limit and and discuss how they change on the addition 
of four derivative terms. In \S\ref{decay}, we discuss the classical decay rate calculation of black holes. We  
review the pole method to compute the scattering cross section, and then apply it to the stringy-corrected 
black holes of interest. In \S\ref{micro}, we discuss the microscopic calculation of the decay rate in the 
two-dimensional conformal field theory describing the system. In \S\ref{discuss}, we end with a discussion 
of our results and of new  calculations that may be interesting to study.

\section{Black holes in string theory} \label{BHreview}

We consider four and five dimensional black holes in type IIB string theory  compactified on
$K3 \times T^{2}$ and $K3 \times S^1$. 
In the five dimensional theory, the black hole carries $Q_{1}$ units of D1-brane charge 
wrapped on $S^{1}$, $Q_{5}$ units of D5-brane charge wrapped on $K3 \times S^{1}$, and $n$ units of momentum on $S^{1}$. 
In the four dimensional theory, there is an additional charge which is a KK monopole corresponding to the 
$\wt S^{1}$ circle, where the $T^{2}$ is represented as $S^{1} \times \wt S^{1}$. 

By string-string duality in six dimensions, the above compactifications to four and five dimensions
can be equivalently described by heterotic string theory on  $T^{4} \times T^{2}$ and $T^{4} \times S^{1}$ 
respectively where the parameters of the $T^{4}$ and the Wilson lines of the heterotic theory are related 
to the parameters of the $K3$ in the IIB description. In the heterotic frame, the five dimensional black hole  has  
$Q_{1}$ units of momentum along $S^{1}$, $Q_{5}$ units of F-string charge wrapped on $S^{1}$, 
and $n$ units of NS5-brane charge wrapped on $T^{4} \times S^{1}$. In the four dimensional theory,
there is in addition, one KK monopole corresponding to the $\wt S^{1}$ circle as before. 

The four and five dimensional black holes described above are related by the $4d$-$5d$ lift \cite{Gaiotto:2005gf}. The 
$5d$ black hole can be obtained by zooming in on the region near the tip of the KK monopole in four dimensions 
where it approximates flat space. 
In particular, the near-horizon geometries of the two black holes are identical,  and since 
the entropy functions depend only on the near-horizon values of the fields  \cite{Sen:2007qy}, the entropy functions 
expressed in terms of the potentials are also identical. However, the actual microscopic values of the entropies 
in terms of the charges differ at subleading order in the large charge expansions ({\it i.e.} at order $\a'$ in the 
macroscopic theory) \cite{Castro:2008ys}. 
In the macroscopic description, this difference arises because of the mixed 
gauge-gravitational Chern-Simons coupling in five dimensions $\int d^{5}x \, A \wedge R \wedge R$ which gives 
a contribution to the right hand side of Maxwell's equation in the four derivative theory \cite{Castro:2007hc}. 
This causes a shift in the definition of the charges in the $4d$ and $5d$ theories and thus to the entropy when 
expressed in terms of the charges, reproducing the $4d$ and $5d$ microscopic results exactly \cite{Castro:2008ys}.

Since this small difference is well-understood both in the microscopic and macroscopic pictures, 
we shall work here with the $4d$ black hole for convenience. We shall express our results in terms of the 
geometric parameters of the $4d$ black hole. To get the final results for the scattering cross section in terms of the charges,
we shall then substitute the values of the parameters in terms of the charges. 
The expression for the cross section for the $5d$ black hole in terms of the geometric parameters is be the same 
(since it is determined by the near-horizon parameters). In order to obtain the corresponding results for the $5d$ black hole, 
we shall substitute the slightly shifted values for the parameters in terms of the $5d$ charges. 
This should become clear in the end of the following subsections.

To compute the corrections to the scattering cross section arising at order $\a'$, we will need the action 
of the theory at four-derivative level. To this end, we use the known complete action of heterotic string theory 
in ten dimensions at order $\a'$ \cite{Metsaev:1987zx}. The theory in four and five dimensions arise upon compactification 
on a $T^{6}$ and $T^{5}$. We shall simply use the dimensional reduction of the ten dimensional action, 
an approximation which is justified in the classical theory. More precisely, the approximation is valid in a 
regime when the string coupling is parametrically small everywhere and the size of the torus is finite in strings units,
both of which shall be true in our solution.

To summarize, we will simply work with the three/four charge solutions to the 
heterotic $10d$ string action. These have been discussed (in the supergravity limit $\a' \to 0$) in \cite{Horowitz:1996ay}, 
\cite{Cvetic:1996gq}. We will discuss the classical solutions in \S\ref{34charge} and the corrections to the near-horizon 
geometry in the extremal limit  in \S\ref{stringy}.

\subsection{The classical three-charge and four-charge black hole} \label{34charge}

We are interested in heterotic string theory on $T^4 \times S^1 \times \tilde{S}^1$. 
The ten-dimensional action for the metric, dilaton and the two form potential\footnote{All the solutions that we shall discuss involve only these fields as we turn off the heterotic gauge fields.} is: 
\be\label{hetact}
S_{\0}=\frac{1}{2 \kappa^2} \int d^{10} x \sqrt{-g} \, e^{-2 \Phi} \, \left( R + 
                       4 \left( \partial \Phi \right)^2 -\frac{1}{2} \left\vert H_{3} \right\vert^2 \right) \,  ,
\ee
where $2 \kappa^2 = (2 \pi)^7 l_s^8 g_s^2$.
The solution\footnote{One way to get this solution is to follow the chain of dualities from type IIB to the heterotic theory described in the 
introduction to this section.
Practically, one can simply use the following chain of steps: first, we S-dualize the type IIB solution: 
\be  \label{IIBtoHet}
G^{\prime}_{\mu \nu}=e^{-\Phi} G_{\mu \nu} \, ,  \qquad
\Phi^{\prime} = -\Phi \, ,  \qquad
B_{(2)}^{\prime} = C_{(2)} \, ,
\ee
thus reaching at the type IIB solution with purely NS-NS flux. This is also a solution of the heterotic action 
which is identical to the type II action in this sector. 
%
} 
for the three-charge black hole is  \cite{Horowitz:1996ay}:
\be\label{hetsol}
\ell_s^{-2} ds_{10}^2  = \frac{1}{\sqrt{h_{1}(r) h_{5}(r)}} \left( -f(r) {dt^\prime}^2 + {dx_5^\prime}^2 \right)
                   + \sqrt{\frac{h_{1}(r)}{h_{5}(r)}} dx_i dx^i
                   + \sqrt{h_{1}(r) h_{5}(r)} \left( \frac{dr^2}{f(r)} + r^2 d \Omega_{S^3}^2 \right) \, ,
\ee
where the thermal and the harmonic functions are given by: 
\be
f(r) = 1-\frac{r_0^2}{r^2}  \, , \quad h_{1,5}(r) = 1 + \frac{r_{1,5}^2}{r^2}  \, .
\ee
Here $(t^\prime, x_5^\prime)$ are the boosted coordinates:
\be
\left( \begin{array}{c}
t^\prime\\
x_5^\prime \end{array} \right) =
\left( \begin{array}{cc}
\cosh \sigma & -\sinh \sigma \\
-\sinh \sigma & -\cosh \sigma  \end{array} \right)
\left( \begin{array}{c}
t\\ x_5 \end{array} \right)  
\ee
and the second term in (\ref{hetsol}) corresponds to $T^4$ with the volume $(2 \pi)^4 V$. 
There is also a non-constant dilaton and the NS-NS 3-form:
\be 
e^{2 \Phi} = \frac{h_{5}(r)}{h_{1}(r)}, \qquad \ell_s^{-2} H_{(3)} = 2 r_5^2 \epsilon_3 + dh_1(r)^{-1} \wedge dr \wedge dx_5.
\ee
Here $\epsilon_3=\frac{1}{8} d\theta \wedge \sin \theta d \phi \wedge d \psi$ 
is the 3-sphere volume form. 

The solution has three independent parameters $r_1$, $r_5$, $r_{0}$ and $\sigma$. 
From the quantized charges of the $H_{(3)}$-form:
\be
Q_5 \equiv  \frac{1}{(2 \pi)^6 \, \ell_{s}^{6}} \int_{S^3 \times T^4} e^{- 2 \Phi} \star_{10} H_{(3)} \qquad \textrm{and} \qquad
n \equiv \frac{1}{(2 \pi)^2 \, \ell_{s}^{2} \, g^{2}_s} \int_{S^3} H_{(3)}
\ee
we find:
\be \label{r1r5}
r_{1}^2 = \frac{Q_{5}}{V} \qquad \textrm{and} \qquad  r_{5}^2 = g_s^{2} n  \,.
\ee
The numbers $Q_{5}$ and $n$ correspond  to the number of fundamental strings and NS5-branes 
respectively\footnote{Note that the names of the integer charges refer to the dual type II system discussed above.}. 
The third charge in the model is the quantized momentum along the compact $x_5$ direction:
\be \label{n}
Q_{1} = \frac{R^2 V}{2 g_s^2} r_0^2 \sinh(2 \sigma).
\ee
Here $2 \pi R$ is the periodicity of the $x_5$ coordinate which is a free parameter of the solution.

Finally, in terms of the parameters $r_{0,1,5}$ and $\sigma$, the effective left and right temperatures are:
\be  \label{TlTr}
T_{L} = \frac{1}{2 \pi} \frac{r_0 e^\sigma}{r_1 r_5}, \qquad
T_{R} = \frac{1}{2 \pi} \frac{r_0 e^{-\sigma}}{r_1 r_5} \, 
\ee
and so the Hawking temperature ($T_{H}^{-1} = \half(T_{L}^{-1}+T_{R}^{-1})$) is:
\be   \label{TH}
T_{H} = \frac{1}{2 \pi} \frac{r_0 }{r_1 r_5 \cosh \sigma}.
\ee
From the form of $T_R$ in (\ref{TlTr}) it is evident that the extremal limit is given by 
$\sigma \to \infty$. Notice though that merely
sending $\sigma$ to infinity, the momentum $n$ in (\ref{n}) also becomes infinite. 
In order to retain a finite $n$ (and $T_L$!) in the extremal limit
we have to simultaneously take $r_0 \to 0$ while keeping the parameter:
\be  \label{rn}
r_n \equiv r_0 \sinh\sigma
\ee
fixed. Notice that $T_H$ still goes to zero as expected.

It is important to address the validity of the supergravity approximation. 
Apart from the small string coupling $g_s \to 0$ requirement we have to ensure that all the scales involved in
the metric \eqref{hetsol} are large compared to the string scale $\ell_s$. This includes the parameters 
$r_1$, $r_5$ and the physical radius of the circle $S^{1}$ everywhere in spacetime\footnote{In the following, 
we will see that in the extremal limit, the radius of the circle near the horizon is $r_{1}$, while at infinity, 
it is equal to $R$. Both of these parameters should be large.}.  
We keep the volume of $T^{4}$ to be finite in string units. 
From \eqref{r1r5} and \eqref{n}, we get\footnote{This can be achieved by scaling:
\be
g_s \sim \lambda^{-1} \, , \quad
n   \sim \lambda^3 \, , \quad
Q_5 \sim \lambda \, , \quad
Q_1 \sim \lambda^{3} \, , \qquad
\textrm{where} \quad \lambda \to \infty \, .
\ee}: 
\be \label{sugraapprox}
g_{s}^{2} n \gg 1 \ , \quad   Q_{5} \gg 1 \ , \quad  g_{s}^{2} Q_{1} \gg 1 \ . 
\ee

Let us now address the geometry described by the metric (\ref{hetsol}) for various ranges of the radial coordinate.
For $r \gg r_{0}, r_{1,5}$, the solution asymptotes to ten dimensional flat space. In the near-horizon region 
$r_{0} \ll r \ll r_{1,5}$, the six-dimensional part of the solution describes a BTZ black hole\footnote{The standard radial coordinate of the BTZ space is given by $\rho^2 = r^2 + r_0^2 \sinh^2 \sigma$.} times a three sphere 
$S^{3}$. Finally, for the ``very-near horizon'' region  
$r \approx r_{0}$, this part of the solution is $AdS_{2} \times S^{1} \times S^{3}$ \cite{Strominger:1998yg}.

We will be interested in evaluating a near-horizon quantity in the extremal $T_H \to 0$ limit. 
As explained in \cite{Sen:2007qy, Faulkner:2009wj,Paulos:2009yk}, this can be achieved by taking the near-horizon limit 
and the extremal limit simultaneously so as to get an ``$AdS_{2}$ black hole''. To be a bit more specific, we
cannot study scattering form the black hole directly in the extremal limit since for $T_H=0$ the horizon is pushed 
to an infinite distance, and the scattering problem is ill defined as such. To remedy this we can follow 
\cite{Sen:2007qy, Faulkner:2009wj,Paulos:2009yk} where we take the near horizon limit and the 
extremal limit simultaneously always keeping the horizon at a finite distance. 
To this end, we introduce new coordinates:
\be\label{extrlim}
r^2 = r_0^2 \frac{z+z_0}{2z_0} \, , \quad t = - z_0 \frac{r_1 r_5}{4 r_n} e^{2\sigma} \cdot \tau \, ,  \quad
x_5 = r_1 \cdot \chi_{4} + z_0 \frac{r_1 r_5}{4 r_n} ( e^{2\sigma} - 2) \cdot \tau \, 
\quad \textrm{and} \quad \psi=2\chi_5 \ ,
\ee
where $\chi_{4,5}$ are periodic with period $2 \pi$.
Taking the $(r_0,\sigma)\to (0, \infty)$ limit keeping $r_n$ in (\ref{rn}) fixed, we arrive at:
\bea\label{vnh}
g_s^{-1} \ell_{s}^{-2} ds^{2} & = & \frac{v_{1}}{16} \left(-(z^{2}-z_{0}^{2})\, d \tau^{2}  + \frac{dz^{2}}{z^{2}-z_{0}^{2}} \right) 
+ \frac{v_{2}}{16}  \left(d \theta^{2}  + \sin^{2} \theta \, d \phi^{2} \right)  + \cr
&& \qquad + \, u_{1}^{2} \left( d \chi_{4} + \half \wt e_{1} (z-z_{0}) d \tau \right)^{2} + u_{2}^{2} \left( d \chi_{5} - \half K  \cos \theta \, d \phi \right)^{2} + dx_{i} dx^{i} \, ,   \cr
e^{-2 \Phi} = \frac{g_s^2}{V} \frac{u_S}{u_1 u_2},   &&  
\ell_s^{-2} H_{(3)} = \frac{1}{2} \frac{v_1 u_1^2}{v_2 u_S} w \cdot d \tau \wedge d z \wedge d \chi_4  +
                    4 \pi \frac{v_2 u_2^2}{v_1 u_S} \tilde{e}_3 \cdot d \theta \wedge \sin \theta d \phi \wedge d \chi_5 \, .
\eea
In terms of the original extremal parameters\footnote{
In \eqref{vnh} we re-introduce $g_s^{-1}$ as an overall metric factor in order to be consistent with the results of 
\cite{Sahoo:2006pm}, where $g_s$ was set to one. As a result, in our notations the parameters 
$v_{1,2}$, $u_{1,2,S}$ and $\tilde{e}_{1,3}$ are $g_s$-independent.} $r_1$, $r_5$ and $r_{n}$:
\be  \label{vuer1r5rn}
v_{1,2} = \frac{4  r_5^2}{g_s}, \quad u_{1} = \frac{r_n}{\sqrt{g_s}}, \quad 
u_{2} = \frac{r_5}{\sqrt{g_s}}, \quad \tilde{e}_{1} = \frac{r_5}{r_n}, \quad 
\tilde{e}_{3} = -\frac{V}{8 \pi g_s} \frac{r_n r_1^2}{r_5}, \quad 
u_{S} = \frac{V}{g_s} \frac{r_n r_1^2}{r_5} \, .
\ee
We have introduced a new parameter $K$, which takes the value $K=1$ for the five dimensional black-hole solution \eqref{hetsol} that we began with. 
\eqref{vnh} is a solution to the heterotic string action \eqref{hetact} for any value of $K$, and it is the near-horizon limit of the near-extremal black hole (in the sense explained above) 
for a black hole carrying four charges when the heterotic
theory is further compactified on a circle $\wt S^{1}$. The charge $K$ as explained 
in the introduction to this section corresponds to the Kaluza-Klein monopole number in going 
from five dimensions to four dimensions. 
The meaning of the other charge $w$ appearing in $H_{(3)}$ 
will be clear in a moment.

Before proceeding, it is worth mentioning that due to the time re-scaling in (\ref{extrlim}), 
the Hawking temperature (\ref{TH}) goes to the limiting finite value:
\be  \label{THvhn}
T_{H} = \frac{z_0}{2 \pi}  \, .
\ee
The temperature of the extremal black hole is of course $T_{H} =0$. The meaning of this finite value 
is that for all scattering processes off the black hole, the physical answers in the extremal limit can 
be computed by considering the above geometry \eqref{vnh} with the effective temperature \eqref{THvhn}. 
The parameter could be thought of as the finite position of the horizon. 
We could have changed the value of the parameter $z_{0}$ by changing the limiting process slightly, 
reflecting the ambiguity of the ``finite'' position of the horizon. In all classical calculations (like all the ones 
we do in this paper), this ambiguity does not affect the answer. 
Indeed the parameter $z_{0}$ eventually drops out of all our physical calculations.

Starting from the general ansatz (\ref{vnh}) one can solve for the free parameters using the  
{\it entropy  function formalism} \cite{Sen:2007qy}.  
The solution can be  summarized  as the minimization of a certain scalar function  $\CE$ of the near-horizon 
parameters, called the entropy function.  This function is constructed so that its value at the extremum is equal 
to the Wald entropy of the black hole in the extremal limit. 

We present the extremal background (\ref{vnh}) using the conventions of \cite{Sen:2005iz}, where
the entropy function for the two derivative theory was calculated.  The same notations were also
adopted in \cite{Sahoo:2006pm} for the calculation of stringy corrections.  Our conventions are different in
two minor aspects which are still worth mentioning. First, the authors \cite{Sen:2005iz,Sahoo:2006pm}
set $\ell_s=4$ while we prefer to keep $\ell_s$ explicitly in the formulae. This explains the $1/16$ factors in
the metric. Second, $\chi_4$ and $\chi_5$ are equal to $x_4/\ell_s$ and $x_5/\ell_s$ of \cite{Sahoo:2006pm}. 
Finally, $z_0=0$ in \cite{Sen:2005iz,Sahoo:2006pm}. Their conclusions, however, are still directly applicable 
also for $z_0 \neq 0$.

With these remarks, we can now cite the entropy function of \cite{Sen:2005iz} for the 
solution\footnote{The charges in \cite{Sen:2005iz} can be identified with ours as 
$(n,w,N^{\prime},W^{\prime}) = (Q_{1}, Q_{5},K,n)$.}
(\ref{vnh}):
\be \label{E0}
\frac{\CE_0}{2 \pi}=
 \frac{Q_{1} \tilde{e}_1}{2}  - 4 \pi n \tilde{e}_3 - \frac{ v_1 v_2 u_S }{4}
 \left( -\frac{1}{v_1} + \frac{1}{v_2} +  \left( \frac{u_1 \tilde{e}_1}{v_1}   \right)^2 +
         \left( 8 \pi \frac{u_2 \tilde{e}_3}{v_1 u_S}  \right)^2 
          -  \left( \frac{u_2 K }{v_2} \right)^2 - \left( \frac{u_1  Q_{5}}{v_2 u_S}\right)^2
\right) \, ,
\ee
Minimizing with respect to $v_1,v_2, u_1, u_2, u_S, \tilde{e}_1$ and $\tilde{e}_3$ we get:
\bea  \label{vueNWnw0}
&& 
v_{1,2}^\0 = 4 K n \, , \quad 
u_{1}^\0 = \left( \frac{Q_{1}}{Q_{5}} \right)^{1/2} \, , \quad 
u_{2}^\0 = \left( \frac{n}{K} \right)^{1/2} \, , \quad 
u_{S}^\0 = \left( \frac{Q_{1} Q_{5}}{K n} \right)^{1/2} \,                    \cr
&& \qquad
\tilde{e}_{1}^\0 = \left( \frac{K n Q_{5}}{Q_{1}} \right)^{1/2} \, , \quad
\tilde{e}_{3}^\0 = - \frac{1}{8 \pi} \left( \frac{K Q_{1} Q_{5}}{n} \right)^{1/2} \, .
\eea
Here the $\0$ subscript reminds us that (\ref{vueNWnw0}) does not include stringy corrections.
As discussed in the introduction to this section, the above near-horizon parameters can be 
related to the $5d$ black hole charges with the simple relation \cite{Dabholkar:2010}, \cite{Castro:2008ys}:
\be\label{4d5drel}
(Q_{1}, Q_{5},n)_{5d} = (Q_{1}, Q_{5},n+1)_{4d} \ . 
\ee
These values which minimize the entropy  function of course agree with the original solution (\ref{vuer1r5rn}).

\subsection{Stringy corrections} \label{stringy}

Stringy effects on low energy physics  can be summarized as higher 
derivative corrections to the Einstein action, suppressed by a corresponding power of the string length.  
Such higher derivative theories have been considered in great detail in the context of corrections to black hole 
entropy (see {\it e.g.} the review \cite{Sen:2007qy}, and in the specific 
heterotic context in  \cite{Sahoo:2006pm}, \cite{Cvitan:2007hu}, \cite{Exirifard:2006qv}). 
We shall consider here the complete four derivative action of string theory in ten dimensions  
\cite{Metsaev:1987zx} dimensionally reduced to four or five dimensions. 

The corrections to the $R^{2}$-type terms that we discuss below come from a four graviton amplitude 
in ten dimensions and are protected by supersymmetry. 
Due to the decoupling of the hypermultiplets and the vector multiplets, these terms 
do not receive any contributions from the moduli of the $T^{6}$ \cite{Harvey:1995fq, Gregori:1997hi}. There {\it are} 
non-trivial heterotic 5-brane instanton contributions to these 
amplitudes, but they can be suppressed by taking the heterotic string coupling to be small
everywhere in spacetime  and the volume of the torus to be 
fixed and finite in string units. From \eqref{vueNWnw0}, we see that  this is achieved by our scaling of the 
charges namely the dilute gas approximation, or the heterotic Cardy limit $1 \ll Kn \ll Q_{1} Q_{5}$. 

A more direct argument which justifies the dimensional reduction is the 
following\footnote{We thank A. Sen for explaining this argument to us.} -- in ten dimensions, 
one can start from the tree level action due to the heterotic coupling being small as mentioned above. 
Since the black hole does not carry any momentum modes along the torus, there is a consistent 
truncation at tree level where all the Kaluza-Klein modes carrying momentum can be set to zero. 
These arguments were given in the 
study of black hole entropy\footnote{In the context of supersymmetric 
black hole entropy, it has often proved to be sufficient to keep 
only a certain class of higher derivative terms (F-type, Gauss-Bonnet etc). These may be justified using 
non-renormalization theorems using or the near-horizon symmetry or supersymmetry. 
Although we will take an extremal limit at the end of our calculations, the process we are studying is intrinsically non-BPS
and it is not {\it a priori} clear that any special type of action captures this physics. It would be interesting to 
understand if and why any such phenomenon happens. }. 
 \cite{Sahoo:2006pm}.

The ten-dimensional heterotic string action to four derivative order is  \cite{Gross:1986mw, Metsaev:1987zx,Hull:1987pc}:
\be \label{hetact4d}
S = S_{\0} + S_{\1}^\prime + S_{\1}^{\prime \prime} \, ,
\ee
where $S_{\0}$ is the two derivative action \eqref{hetact} and: 
\begin{eqnarray}  \label{defS1}
S_{\1}^\prime &=& \frac{1}{2 \kappa^2}\, \frac{\alpha'}{8}\, \int d^{10} x 
\sqrt{-\det {g}}~e^{-2 \Phi}\,\left( { R_{klmn} R^{klmn}} -\frac{1}{2} R_{klmn} H_{p}^{~kl} H^{pmn} \,\right.
\cr
&&  
\left.\qquad -\frac{1}{8} H_{k}^{~mn} H_{lmn} H^{kpq} H^{l}_{~pq}
 + \frac{1}{24} H_{klm} H^{k}_{~pq} H_{r}^{~lp} H^{rmq}\right).
\end{eqnarray}
The last term $S_{\1}^{\prime \prime}$ in (\ref{hetact4d}) is due to the fact that in the $\alpha^\prime$-corrected theory the
NS-NS $H_{(3)}$ form is not closed anymore:
\be  \label{HdBOmega}
H_{(3)} = d B_{(2)} + 3 \alpha^\prime \Omega_{(3)} \, .
\ee
Here $\Omega_{(3)}$ is the gravitational Chern-Simons 3-form satisfying:
\be
d \Omega_{(3)} = - \frac{1}{8} R_{(2)} \wedge R_{(2)} \, .
\ee
Substituting (\ref{HdBOmega}) into \eqref{hetact} one finds an additional contribution $S_{\1}^{\prime \prime}$
to the four-derivative action. 

Once we turn on stringy corrections, \eqref{hetsol}--\eqref{r1r5} is no longer a solution. We expect to find an asymptotically flat 
solution at any order in $\a'$. In general, a full analytic solution has not been found, but there is numerical evidence for the existence of such a black hole solution \cite{Castro:2007hc}. 
However, for our purposes, we will need only the very near-horizon solution
in the extremal limit  which always has the form \eqref{vnh} above since this is the most general solution consistent 
with the emergent isometries in this limit \cite{Kunduri:2007vf}. The values of the parameters $v_{1}, v_{2}$,  \emph{etc} will of course receive corrections compared to their values in the two derivative theory (\ref{vueNWnw0}), {\it e.g.} we would not expect $v_{1} = v_{2}$ anymore.

Once we restrict our attention to the very near horizon solution (\ref{vnh}) we can use the results of 
\cite{Sahoo:2006pm} where the entropy function for the Ansatz (\ref{vnh}) and the $\alpha^\prime$-corrected
action (\ref{hetact4d}) were calculated for the first time perfectly matching the microscopical expectations. As was pointed out in
\cite{Sahoo:2006pm} one cannot apply directly the Wald formula 
(and as a consequence the entropy function formalism) to the action (\ref{hetact4d}) because the Chern-Simons
3-form cannot be rewritten in a manifestly covariant form (in other words it doesn't depend on the Riemann tensor 
explicitly).  The way to deal with the Chern-Simons term was explained in \cite{Sahoo:2006vz}.
One should first introduce a new 3-form $\mathcal{K}_{(3)}$ which serves as a Lagrange multiplier for
(\ref {HdBOmega}) and then dimensionally reduce the action to four dimensions. A new 4$d$ action
obtained this way is manifestly covariant and the application of the entropy function formalism 
is now straightforward\footnote{The same result for the BTZ black hole entropy was later derived in \cite{Tachikawa:2006sz} 
based only on Wald's Noether charge argument.}.

For the four-derivative action \eqref{hetact4d} the entropy function is \cite{Sahoo:2006pm}:
\be  \label{Etotal}
\CE = \CE_{0} + \CE_1^\prime + \CE_1^{\prime \prime} \, ,
\ee
where $\CE_{0}$ is given in (\ref{E0}), $\CE_1^\prime$ corresponds to the (\ref{defS1}) contribution and
$ \CE_1^{\prime \prime}$ takes into account the Chern-Simons term due to (\ref{hetact4d}).
It was found in \cite{Sahoo:2006pm} that:
\begin{eqnarray}  \label{E1prime}
\frac{\CE_1^\prime}{2 \pi} &=& -2 v_1 v_2 u_S \left[ \frac{1}{2 v_1^2} + \frac{1}{2 v_2^2} 
                                      - 3 \frac{u_1^2 \tilde{e}_1^2}{v_1^3} - 3 \frac{u_2^2 }{v_2^3} {K}^2
                                  + \frac{11}{2} \frac{u_1^4 \tilde{e}_1^4}{v_1^4} + \frac{11}{2} \frac{u_2^4}{v_2^4} {K}^2 
                                   - \frac{u_1^2}{v_1 v_2^2 u_S^2} Q_{5}^2 - \right.
\cr
&&  \left.
    - 64 \pi^2 \frac{u_2^2 \tilde{e}_3^2}{v_1^2 v_2 u_S^2} 
    + \frac{u_1^4 \tilde{e}_1^2}{v_1^2 v_2^2 u_S^2} Q_{5}^2 
    + 64 \pi^2 \frac{u_2^4 \tilde{e}_3^2}{v_1^2 v_2^2 u_S^2} {K}^2
    - 10 \frac{u_1^4}{v_2^4 u_S^4} Q_{5}^2 - 10240 \pi^4 \frac{u_2^4 \tilde{e}_3^4}{v_1^4 u_S^4}  \right] 
\end{eqnarray}
and:
\be \label{E1primeprime}
\frac{\CE_1^{\prime \prime}}{2 \pi} = 
                    4 \left( 1 - 2 \frac{u_1^2}{v_1} \tilde{e}_1^2 \right) \frac{u_1^2}{v_1} \tilde{e}_1 Q_{5} 
								- 32 \pi \left( 1 - 2 \frac{u_2^2}{v_2} {K}^2 \right) 
                                                                \frac{u_2^2}{v_2} \tilde{e}_3 K
\, .
\ee
To find the entropy form the entropy function it suffices to find the values that minimize
(\ref{E0}) and plug it into (\ref{Etotal}). Indeed, the $\alpha^\prime$-corrections to the minimizing solution of
(\ref{Etotal}) will contribute only at the order ${\alpha^\prime}^2$ once the solution is plugged back into (\ref{Etotal}),
and therefore these corrections can be neglected in the entropy calculation.
For our goals in the next section, however, we will need the $\alpha^\prime$-corrections of $v_1$, $v_2$ and 
some other parameters. A straightforward analysis shows that the $\alpha^\prime$-expansion parameter 
for our dimensionless parameters is $(K n)^{-1}$. The simplest way to see this is to notice that 
according to (\ref{vueNWnw0})
at the leading order the curvature of both the $AdS_2$ and the $S^2$ in (\ref{vnh}) goes as 
$v_{1,2}^{-1} \sim (K n)^{-1}$ and thus the curvature $\alpha^\prime$-corrections should be
supressed by the $(K n)^{-1}$ factor.
Let us denote by $v_1^\1$ the first subleading term in the $\alpha^\prime$-expansion of $v_1$:
\be
v_1 = v_1^\0 + v_1^\1
\ee
and similarly for the other parameters $v_2$, $u_S$, $u_1$, $u_2$, $\tilde{e}_1$  and $\tilde{e}_3$.
One finds that the following set minimizes (\ref{Etotal}) \cite{Cvitan:2007hu}:
\begin{eqnarray}  \label{vueNWnw1}
&&
v_1^\1 = 4 \, ,  \quad
v_2^\1 = 12 \, , \quad
u_1^\1 = - \frac{3}{2  K n } \sqrt{ \frac{Q_{1}}{Q_{5}} } \, , \quad
u_2^\1 = \frac{3}{2  K n }  \sqrt{\frac{n}{K }} \, , \quad
\cr
&&
u_S^\1 = -\frac{2}{K n} \sqrt{ \frac{Q_{1} Q_{5}}{ K n } } \, , \quad    
\tilde{e}_1^\1 = \frac{2}{\sqrt{K n} } \sqrt{ \frac{Q_{5}}{ Q_{1} } } \, ,  \quad
\tilde{e}_3^\1 = \frac{1}{4 \pi K n} \sqrt{ \frac{K  Q_{1} Q_{5}}{ n } } \, .
\end{eqnarray}

\section{The scattering cross section and decay rate} \label{decay}

\subsection{The classical result} \label{sigmaclass}

The scalar plane wave scattering in the (near)-extremal version of type IIB background dual to the heterotic solution 
of the previous section was thoroughly analyzed in \cite{Dhar:1996vu} and \cite{Das:1996wn}, following the recipe 
of Unruh \cite{Unruh:1976fm}. As we mentioned in 
the introduction, for very low energies the total absorption cross section is equal to the horizon area, as consistent with the 
microscopic calculation. To arrive at this result one has to consider a minimally coupled scalar in this background, 
which is the most general form at low energies for any linear scalar perturbation. Since an analytic solution for 
the emerging Klein-Gordon equation is not known for the entire range of the radial coordinate, one considers  
three different regimes and then glues the corresponding solutions in a smooth way. 

In the outmost region $r \gg r_{1,5}$, the metric essentially becomes flat and so the solution is given by a 
superposition of an incoming and an outgoing spherical wave\footnote{These $s$--waves are described by Bessel functions.}. 
As we move into the $r \approx r_{1,5}$ region the two independent solutions are $r^{-2}$ and a constant. 
Once we approach the horizon, however, the Schwarzschild coordinates become singular and we have to switch 
instead to the non-singular Eddington-Finkelstein coordinates\footnote{The new radial coordinate is also called the 
tortoise coordinate.}, about which we shall say more in the next subsection. 
In the new coordinates the Klein-Gordon equation once again yields an infalling and an outgoing solution.  
In order to find the \emph{absorption} cross section we should use the former. 

Now, starting from the chosen solution near the horizon, we can move outwards and glue the solutions on 
the different patches.  In this manner, one computes the ratio between the incoming and the outgoing waves 
in the asymptotic infinity. Knowing this ratio, one can calculate the absorption probability. This is not yet the 
final answer though, since the microscopic calculation is performed with a plane wave at infinity.
Rewriting the $s$-wave in terms the plane wave, the authors of \cite{Dhar:1996vu,Das:1996wn} found the 
scattering cross section which is exactly the horizon area. This result was later generalized in \cite{Das:1996we} 
to low energy scattering in an arbitrary black hole background, with exactly the same conclusion -- the low energy
cross section is universally equal to the horizon area.

\subsection{Scattering cross sections and the pole method} \label{pole}

In the above subsection, we described the conventional calculation of the cross section as a ratio of fluxes falling 
on the black hole in a general \emph{two} derivative theory\footnote{There are other ways to define the  cross section which are equivalent to this one \cite{Harmark:2007jy}.}. 
Unfortunately that for higher derivative theories, it is not clear how to define the scattering process from asymptotically 
flat space since the plane waves are not the only solutions\footnote{ One way to overcome this is to use the physical 
notion that the new solutions are suppressed by $\a'$, and are therefore spurious \cite{Sen:2004dp}, \cite{Hubeny:2004ji}. 
}.
The approach we shall use here computes the low-energy cross section 
by evaluating the on-shell action and its derivative on a certain slice of spacetime. 
Similar ideas have been discussed in \cite{Polchinski:1999ry},  \cite{Minwalla:1999xi}. 
The advantage of this procedure is that it naturally generalizes to a theory of higher 
derivatives. The final expression for the cross section for a scalar field $\psi$ scattering off a black hole can be summarized in 
a very simple formula in the spirit of \cite{Iqbal:2008by,Paulos:2009yk},  that we shall  describe in this subsection.

Roughly speaking, the idea is as follows -- using the optical theorem, one relates the scattering cross section to the imaginary 
part of the forward scattering amplitude in flat space. 
This amplitude can be then thought of as a limit of a certain retarded Green's function computed on a radial slice in spacetime. 
This Green's function is defined in one less 
dimension and in this sense holographic\footnote{In a truly holographic context like the fluid-gravity correspondence,
this is really a Green's function of a certain operator living on the boundary of $AdS$, which is related to the 
transport coefficients appearing in the Kubo formula for the fluid living on the boundary \cite{Policastro:2001yc}, 
\cite{Son:2007vk}. 
In our case we do not have any precise notion of a boundary theory since we are in asymptotically flat space. }. 
To match this Green's function with the bulk scattering amplitude,
one solves for the radial momentum of the field using its mass-shell condition.

Remarkably, at low energies the cross section calculated this way is independent of the value of the radial coordinate 
where one evaluates the Green function \cite{Iqbal:2008by} and, in particular, it can be evaluated at the horizon\footnote{
In this sense the method now appears to be deeply connected to the black hole membrane paradigm.}.  
Following the proposal of \cite{Paulos:2009yk} we will adhere to this definition of cross section also for 
higher derivative cases in which case as well there is no radial flow. The only difference is that the radial momentum 
should now be calculated for a Lagrangian with higher derivatives of the fluctuation field. 
This proposal is based solely on the field regularity at the horizon as we shall sketch below.

One assumes the general form of the metric to be \cite{Paulos:2009yk}:
\bea\label{genmetric}
ds^{2} & = & g_{ab} \, dx^{a} dx^{b} = \frac{L^{2}}{z-z_{0}} e^{2g(z)} dz^{2} + g_{\mu\nu} dx^{\mu} dx^{\nu} \ , \cr 
g_{\mu\nu} & = & - (z - z_{0}) e^{2f(z)} dt^{2} + e^{2\rho(z)} dx_{i} dx^{i}   \, ,
\eea
where the functions $f(z)$, $g(z)$ and $\rho(z)$ are regular for any $z \geqslant z_0$. 
The horizon is at $z=z_{0}$, the boundary is at $z=z_{\rm bdry}$ and the Hawking temperature is:
\be
T_{H} = \frac{1}{4 \pi L} e^{f(z_0)-g(z_0)} \, .
\ee
It is not difficult to see that 
the near horizon limit of the metric \eqref{hetsol} is very similar to \eqref{genmetric}.
It is not quite of this form as such, since there is an additional sphere factor. However,
this sphere is perfectly regular at the horizon and will not affect our conclusions\footnote{Strictly speaking, we 
can compactify on this sphere and compute the cross section ``per unit area'', but we shall quote the final 
results after multiplying it back. In practice, we can simply work with the ten-dimensional metric.}. 

We are interested in a scattering process due to a scalar perturbation $\psi$ of the metric. To
study the scattering we have to expand the metric perturbation only to quadratic order. 
In the next section we shall explain precisely which fluctuation we are to consider. 
For the moment it will be enough to notice that up to two derivatives
any such perturbation leads to the action of a minimally coupled scalar field:
\be \label{S2psi}
S^{\2}_{\psi} = - \frac{1}{4 \kappa} \int d^{n} x \, d t \, d z \sqrt{-g} \left( \nabla \psi \right)^2 \, .
\ee
Here $z$ is the radial coordinate, $d^{n} x$ corresponds to the rest of the coordinates not including 
time\footnote{The perturbation scalar $\psi(z,x)$ is related to the scalar field $\phi(r,x)$ of \cite{Iqbal:2008by,Paulos:2009yk} by
$\psi=\phi/\sqrt{2}$. This latter re-scaling is introduced in order to match the formulae of this subsection
to the particular fluctuation we are interested in (to be a bit more specific, the $2 \kappa$ factor in \eqref{S2psi} comes
from the Einstein action coefficient in the supergravity action, while the remaining $2$ factor comes from the $\psi$-expansion).}.

Next, let us ignore for simplicity the spatial dependence of $\psi$ and denote by $\psi_\omega(z)$ the Fourier transform of
$\psi(z,x)$ with respect to all coordinates but the radial one:
\be
\psi(z,x) = \int  \frac{ d^n k \, d t}{( 2 \pi)^{n+1}} \psi_\omega(r) \delta(\mathbf{k}) e^{k_\mu x^\mu}  \, , 
\ee
where $k_\mu = \left(\omega, \mathbf{k} \right)$. 
The formula for the two derivative theory follows from the optical theorem:
\be \label{sigmapsi}
\s_{\rm abs} = 16 \pi G_N  \lim_{z \to z_{\rm bdry}}  \lim_{\omega \to 0} 
            \textrm{Im} \frac{\pi_{\psi} (\omega ,z)}{\omega \psi(\omega,z)} \ , 
 \ee
where: 
\be\label{defpi}
\pi_{\psi}(\w,z) = \frac{\delta S_{\psi}}{\delta (\pa_{z} \psi(-\w,z))}
\ee
is the radial conjugate momentum to the field $\psi$, and 
$z$ is the location of the radial slice which should be taken to the asymptotic boundary 
value $z_{\rm bdry}$ where one has flat Minkowski space. 

One nice feature of the \emph{low energy} limit that we consider is that the quantity related to the 
aforementioned ``holographic" 
retarded Green function:
\be\label{noevolu}
 \lim_{\omega \to 0} \textrm{Im} \frac{\pi_{\psi} (\omega ,z)}{\omega \psi(\omega,z)}
\ee
does not have any radial evolution \cite{Iqbal:2008by} and we can therefore evaluate the right hand side of 
\eqref{sigmapsi} at any value of the radial coordinate $z$, and in particular at the horizon $z=z_{0}$:
\be \label{sigmaformula}
\s_{\rm abs} = 16 \pi G_N  \lim_{z \to z_{0}}  \lim_{\omega \to 0} 
            \textrm{Im} \frac{\pi_{\psi} (\omega ,z)}{\omega \psi(\omega,z)} \ .
\ee
The expression on the right hand side can be checked to be proportional to the horizon 
area $ A_{H} = \left. \int d^n x \right\vert_{z=z_0}$, and we make sure that the 
constant above is correct  by demanding that the expression is equal to the area
to match the known flat space result.

For a general theory with higher derivative $\alpha^\prime$-corrections to \eqref{S2psi}, 
one does not {\it a priori} have a universal form of the action like in \eqref{S2psi}. 
However, any higher derivative \emph{effective} action can always 
be re-written {\it near the horizon} as \eqref{S2psi} with $\kappa$ replaced by an \emph{effective} $\tilde{\kappa}$ \cite{Paulos:2009yk}.  
The derivation is based only on the regularity of the action (and its solution) near the horizon.  
As we already mentioned in the previous subsection, the regularity of
the field $\psi(z,t)$ near the horizon implies that it necessarily depends only on one of the two Eddington-Finkelstein coordinates,
the infalling or the outgoing:
\be  \label{dudv=0}
\left. \frac{\partial}{\partial v} \frac{\partial}{\partial u} \psi(z,t) \right\vert_{\rm horizon} = 0
\, , \quad \textrm{where} \quad
d v \equiv d t + \sqrt{-\frac{g_{z z}}{g_{t t}}} d r
\quad \textrm{and} \quad
d u \equiv d t - \sqrt{-\frac{g_{z z}}{g_{t t}}} d r  \, .
\ee


Remarkably, for this statement to hold, the only condition on the metric is that $g_{z z }$ diverges as $(z-z_0)^{-1}$, 
$g_{t t }$ vanishes like $(z-z_0)$, while all other metric components are regular near the horizon at $z=z_0$.
In terms of $\psi_\omega(z)$ the second order differential equation 
in \eqref{dudv=0} can be recast as:
\be  \label{FTdudv=0}
\left. \left[ \psi^{\prime \prime}_\omega (z) + \frac{\psi^{\prime}_\omega (z)}{z-z_0} + 
           \frac{\omega^2}{\left( 4 \pi T_{H} \right)^2}  \frac{\psi_\omega (z)}{\left( z-z_0 \right)^2} \right]
\right\vert_{z=z_0} 
= 0 \, ,
\ee
where we used the definition of the Hawking temperature in terms of $g_{z z}$ and $g_{t t}$ near the horizon.
This equation in turn is solved by:
\be  \label{dudv=0Solution}
\psi_{\omega}(z) \approx \psi_{0} \exp{\left( \pm i \frac{\omega}{4 \pi T_{H}} \log(z-z_0) \right)}  \, .
\ee
Here the $\pm$ corresponds to the infalling and the outgoing solutions.

Following \cite{Paulos:2009yk} we can now write the most general effective action that 
reproduces \eqref{FTdudv=0} near the horizon as its equation of motion (EOM):
\be  \label{FTS2psi}
S^{\2}_{\rm eff} \approx \frac{1}{ 4 \tilde{\kappa}} \int d^n x \int \frac{d \omega}{2 \pi} \int_{z_0} d z 
    \sqrt{-g} \left( g^{z z} \psi_{\omega}^\prime (z) \psi_{-\omega}^\prime (z) + 
                      g^{t t} \omega^2 \psi_{\omega}(z) \psi_{-\omega} (z)  \right)
        \quad + \quad \ldots \,             
\ee
for some constant $\tilde{\kappa}$.
The full Lagrangian of course does 
contain other higher derivative terms. In practice, one can explicitly 
integrate out the higher order terms in a consistent manner and bring the effective Lagrangian 
to the above form upto boundary terms. These possible boundary terms  are denoted in \eqref{FTS2psi} by the $\ldots$ .

In general, derivation of the effective two-derivative action from the full higher derivative action might be 
a formidable task. It requires integrating out all modes except for the two lowest ones, 
as, for example, was done in \cite{Banerjee:2009wg}.
Astonishingly, according to
the observations of \cite{Paulos:2009yk}, all we need to know for our purposes is only the form
of the effective action near the horizon, which is unambiguously (up to a choice of $\tilde{\kappa}$) given by  \eqref{FTS2psi}.
This is the essence of the \emph{pole method} of \cite{Paulos:2009yk} that we will very
briefly describe in the remainder of this subsection.

For the two-derivative case we have $\kappa=\tilde{\kappa}$ and the action  \eqref{FTS2psi}
is actually equivalent to \eqref{S2psi} near the horizon. In this case, using 
 \eqref{FTS2psi}, we can explicitly check that the cross section equals the area of the horizon. 
For the general case, the  general form of the action  \eqref{FTS2psi} instructs us to 
simply replace $\pi_{\psi}$ in \eqref{sigmaformula} by the generalized conjugate 
momentum $\Pi_{\psi}$ which can depend on higher derivatives as well \cite{Paulos:2009yk}. 
This has the same formal definition as in the two derivative case:
\be\label{defPi}
\Pi_{\psi}(\w,z) = \frac{\delta S_{\psi}}{\delta (\pa_{z} \psi(-\w,z))} \ , 
\ee
but one now has to be careful about potential ambiguities arising from boundary terms. 
This issue was addressed by \cite{Paulos:2009yk}, and the final prescription 
is that 
one only keeps terms in the action quadratic in $\psi$ and such that $\psi$ always appears differentiated. 
Importantly, the quantity \eqref{noevolu} with $\pi_\psi$ being replaced $\Pi_\psi$ is still independent
of the radial coordinate.
Putting all these pieces together, 
one can present the final formula in many different ways\footnote{We have not rigorously derived these formulas 
from the conventional definition of the scattering amplitudes, we postpone a careful analysis 
of this problem for the future \cite{Kuperstein:2010pw}. }. 
The two which we use
are called in \cite{Paulos:2009yk} the \emph{time pole formula}:
\be \label{Pole1}
\frac{\sigma_{\rm abs} } {16 \pi G_N} = -   8 \pi \, T_H  \lim_{\w \to 0} \frac{\displaystyle{\Res_{z \to z_{0}} } \, 
            \d \CL|_{\psi(\tau) = e^{-i \w \tau}}}{\w^{2}} 
\ee
and the \emph{radial pole formula}: 
\be \label{Pole2}
\frac{\sigma_{\rm abs} } {16 \pi G_N}  = 8 \pi \, T_H  \lim_{\w \to 0} \frac{\displaystyle{\Res_{z \to z_{0}} } \, 
   \d \CL|_{\psi(z) = e^{i \frac{\w}{4 \pi T_{H}} \log (z-z_0) }}}{\w^{2}}  \, .
\ee
The derivation of these formulae relies on the solution \eqref{dudv=0Solution} and the fact that in terms of 
the generalized radial momentum $\Pi_\psi$ the effective action reads:
\be
S^{\2}_{\rm eff} = \frac{1}{ 2} \int d z  \, \Pi_\psi(z) \psi^\prime_{\omega}(z)   \, .
\ee

\subsection{Stringy corrections to the cross section} \label{sigmacorr}

We want to apply the method discussed in the previous section for the four-derivative action
(\ref{hetact4d}). In this paper we are interested only in a scalar perturbation that comes from
the metric perturbation along the flat $T^4$ directions. To be more precise, denoting the torus 
directions in the metric by $i=6 \ldots 9$, our 
plane wave scattering process is due to the following fluctuation:
\be  \label{g67}
\delta g_{67} = \psi(r,t). 
\ee
To apply the pole formula (\ref{Pole1}) we have to substitute (\ref{g67}) into the full 
four-derivative action (\ref{hetact4d}) and to expand it to the second order in $\psi(r,t)$ around the background
(\ref{vnh}). 
This will produce plenty of terms. It is worth remembering though, that terms with no 
time or radial derivatives (depending on what formula we choose in (\ref{Pole1})),  of $\psi$
cannot contribute to (\ref{Pole1}). This leaves only the Riemann tensor terms or the CS term contribution 
$S_{\1}^{\prime \prime}$.  The latter does not contribute as can be shown by a straightforward and a slightly tedious
calculation. Moreover, among the three terms depending on the Riemann tensor in (\ref{hetact4d}) the second term
in (\ref{defS1}) also does not contribute since the $H_{(3)}$ form has no legs along the $6,7$ directions. To
sum up, we have to expand only the $R$ and $R^2=R_{k l m n} R^{k l m n}$ terms 
from (\ref{hetact}) and (\ref{hetact4d}) respectively:
\be\label{deltaS}
\delta S  =  
 \frac{1}{2 \kappa^2}\, \int d^{10} x  \sqrt{-\det{g}} \, e^{-2\Phi}  \left(\delta R + \frac{\a'}{8} \delta R^{2} \right) \quad +  
\quad \textrm{irrelevant terms} \, .
\ee
Although by construction the time and the radial formulae (\ref{Pole1}) eventually lead to the same results,
we prefered to use the time version because it slightly shortens the computations..
For $\psi=\psi(\tau)$ we have:
\be\label{corrR}
\ell_s^2 \, \delta R  =     \frac{16}{v_{1}} \cdot \frac{ \partial^2_\tau \left( \psi^2 \right) 
                                        - \frac{1}{2} (\partial_\tau \psi )^2 }{ (z^{2} - z_{0}^{2})}  + \ldots \, , 
\ee
and:
\be\label{corrR2}
\ell_s^4 \, \delta R^{2} =  \frac{1024}{v_1^2} \cdot   \left[
                         \frac{ \frac{1}{2} \partial^2_\tau \psi  
                                        -  z^2 (\partial_\tau \psi )^2 }{ (z^{2} - z_{0}^{2})^2} 
              + \frac{u_1^2 \tilde{e}_1^2}{v_1} \cdot \frac{ (\partial_\tau \psi )^2 }{ (z^{2} - z_{0}^{2})}   \right] + \ldots\, ,
\ee
where, again, the $\ldots$ stands for the terms with no derivatives.
Dropping the total time derivatives and using (\ref{THvhn}) we arrive at:
\be \label{polecalcs}
8 \pi T_H  \lim_{\w \to 0} \frac{\displaystyle{\Res_{z \to z_{0}} } \, \d R}{\w^{2}}  = - \frac{16}{v_1 \ell_s^2} 
\quad \textrm{and}  \quad
8 \pi T_H  \lim_{\w \to 0} \frac{\displaystyle{\Res_{z \to z_{0}} } \, \d R^2}{\w^{2}}  = - \frac{1024}{v_1 \ell_s^4}
                                                            \left( 1  - 2 \left( \frac{u_1 \tilde{e}_1}{v_1} \right)^2 \right)  \, .
\ee
We are now in a position to calculate the total cross section. Using (\ref{Pole1}), (\ref{deltaS}) and (\ref{polecalcs}) we 
find:
\be \label{TCS1}
\sigma_{\rm abs} = 
   2 \pi G_{N} v_2 u_S \left( 1 + \frac{8}{v_1} \left( 1 - 2 \frac{ \left( u_1 \tilde{e}_1 \right)^2}{v_1}  \right) \right) \, .
\ee
In the above expression, we should now substitute the values of the various fields at first and second order derived above in 
(\ref{vueNWnw0}) and (\ref{vueNWnw1}).
The scattering cross section then can be written as:
\be \label{TCS2}
\s^{4d}_{\rm abs}  = 8 \pi  G_{N} \sqrt{Q_{1} Q_{5} K n } \left( 1 + \frac{2}{K n} \right) \, . 
\ee

All this was for the four dimensional black hole. For the five dimensional case, as was discussed in section \S\ref{34charge}, 
we simply need to use \eqref{4d5drel} to derive:
\be \label{TCS25d}
\s^{5d}_{\rm abs}  = 8 \pi  G_{N} \sqrt{Q_{1} Q_{5} n } \left( 1 + \frac{3}{2 n} \right) \, . 
\ee

As was already advertised in the introduction these results are precisely four times the entropy 
and in the next section we will re-derive this result in the microscopic D-brane picture.

\section{The microscopic decay rate} \label{micro}

In this section, we shall compute the rate of emission of a low energy scalar from a near-extremal system 
of branes,  to sub-leading order  in the large charge approximation. In particular, we will consider the 
D1-D5-P system in IIB string theory in five dimensions, and the related four charge system in the four 
dimensional theory obtained by the addition of a KK-monopole. 

The leading order calculation \cite{Das:1996wn} was done in the canonical ensemble, but in the infinite 
charge limit, the choice of statistical ensemble is of course immaterial. 
To go beyond the leading order estimate and to consider finite charge effects, we first need to decide a 
choice of statistical ensemble, which we do in \S\ref{ensemble}. In \S\ref{frames}, we shall discuss the system 
under consideration in different duality frames, and derive thermodynamic relations in each frame. 
Having done this, we find that  the rest of the calculation essentially follows the leading order calculation of 
\cite{Das:1996wn} very closely, which we then sketch in \S\ref{microcorrections}.

\subsection{Choice of ensemble in the near-extremal system} \label{ensemble}

The leading order calculation was originally done in \cite{Das:1996wn} by considering the thermodynamics of the gas of  
open string modes living on the D-branes. The decay process is that of open strings annihilating to form a 
closed string which runs off to infinity. At low energies, this is dominated by two massless open strings 
annihilating to form a graviton. One then has to average over initial states and sum over final states. 

A more modern approach is to consider the two dimensional field theory describing 
the D-brane system coupled to gravity in asymptotically flat space.  At low energies, one can ignore the 
massive stringy modes on the D-branes, and one gets a two dimensional SCFT on the effective string. 
This effective string wraps a circle of radius $R$ which provides an external scale to the theory. 
One then considers the amplitude of the coupling of various vertex operators in this SCFT to the external 
graviton, and then averages over various vertex operators \cite{Avery:2009tu}. 

There are many subtle issues in this approach which arise from the fact that the energy spectrum on the effective 
string is discrete \cite{Das:2008ka}. An obvious issue is that because of the mass gap in the spectrum, 
an external bulk mode with very low energies will not couple to the string. However, if we take the limit 
$R/\ell_{s} \gg1$, the spectrum is well approximated by a continuum, and there is a non-zero coupling of the 
bulk mode even in the extreme low-energy limit. In this limit, there are two other nice features that we obtain. Firstly, 
we can restrict our attention to the lowest energy modes on the effective string {\it i.e.} the ``naive'' coupling 
of the D-branes to the external theory by the Born-Infeld action (or equivalently a disk amplitude with one 
closed string insertion).  Secondly, one can approximate all the sums appearing in the average by integrals. 
To summarize,  we use the setup used in the leading order calculations \cite{Das:1996wn, Maldacena:1997cg} 
(in which these approximations were implicitly or explicitly made). 
Essentially, one can treat the problem as a $1+1$ dimensional gas of massless particles carrying energy and 
momentum, and the coupling given by the geometric embedding of the D-branes in spacetime.

The system we consider is excited slightly above extremality. To model the system as a 
statistical system, we think of 
the supersymmetric ground states of the string as the reservoir with energy = momentum = $k/R$,
with $k \gg1$. The physical energy measured in string units is $E \equiv k\ell_{s}/R$. 
On top of this, we excite the system with energy
$\w \ll E$. By standard 
statistical arguments, the probability $P(\w)$  that the full system exists in the state $E+ \w$ obeys 
the relation:
\be\label{statmech}
\frac{P(\w_{1})}{P(\w_{2})} = \exp(S(E+\w_{1}) - S(E+\w_{2}) ) \ ,
\ee
where $S(E) = \log \O (E)$ with  $\O (E)$ being the exact degeneracy of states of the system.

To compute the statistical average of any quantity using the distribution $P(\w)$, we do a Taylor 
expansion around $E$:
\be\label{Ptaylor}
P(\w) =  K \exp \left( S(E) +\left. \frac{\pa S(E+\w)}{\pa \w} \right|_{\w =0} \cdot \w + \cdots \right)  \, ,
\ee
where $K$ is a fixed constant of proportionality and the higher terms are suppressed by higher powers 
of $\w$. It is clear then that if we are only 
interested in computations in the limit $\o \to 0$, all quantities can be computed by 
keeping only the first term in the above expansion. This means that in the extreme low energy limit,
the exact distribution function can be simply replaced by the 
the {\it canonical ensemble} with fixed temperature given by:
\be\label{defT}
\left. \frac{1}{T} \equiv \frac{\pa S(E+\w)}{\pa \w} \right|_{\w=0} \ .
\ee
With this temperature, one can use the canonical distribution function: 
\be \label{defrho}
\rho(E) = \frac{1}{e^{E/T} \mp 1}  \, ,
\ee
where the $-$ sign is for bosons and the $+$ sign is for fermions.

\subsection{The conformal field theory in different duality frames} \label{frames}

Most examples of black holes in string theory 
descend from an effective black string wrapped around a circle
in a string compactification. An effective description of this system is as a string 
wrapping a circle $S^{1}$ of radius $R$ $w$ times. The mass gap on a string wrapping the circle is naively is $1/R$, but 
due to the winding, it is reduced to $1/wR$. This is interpreted as the lowest momentum mode on a {\it long string}  
that has an effective length $L=2 \pi w R$ \cite{Maldacena:1996ds}.  

On this long string, there is an effective conformal field theory with central charge $c_{{\rm eff}}$ 
which is typically small, it describes the geometric motion of one long string in the internal spacetime.  
If we consider the thermodynamics of the $1+1$-dimensional gas of massless modes on the long string, 
we find quite generally the entropy-temperature relation \cite{Das:1996wn}: 
\be\label{STrel1dgas}
S =  \frac{c_{{\rm eff}}}{6} \pi L T \ . 
\ee

We will now be more precise and describe our system of interest. 
As discussed in the previous sections, there are different dual descriptions of the system we consider. 
The best-understood one in the microscopic setting is the IIB theory compactified on $K3 \times S^{1}$,
with D1-D5-P charges. 
The dual heterotic description involves a compactification on $S^{1} \times T^{4}$ with P-F1-NS5 charges. 
Due to the presence of the NS5-brane, the low energy limit is not as well understood {\it per se}. 
One can go to four dimensions by compactifying on another circle $\wt S^{1}$ an adding a Taub-NUT charge 
associated with this circle. 

In the type IIB frame, the supersymmetric configuration 
consists of D5-branes wrapping $K3 \times S^{1}$, D1-branes wrapping $S^{1}$,   and 
momentum on $S^{1}$ with net charges $(Q_{1},Q_{5},n)$ as considered in the 
macroscopic calculation. All the momentum 
is purely left-moving to ensure supersymmetry. In the extreme low energy limit, one 
recovers a two-dimensional superconformal field theory with target space: 
\be\label{D1D5}
{\rm (Sym)}^{Q_{1}Q_{5} +1} (K3) \times \IR^{4} \ ,
\ee
with the Hamiltonian $L_{0}=n$ \cite{Strominger:1996sh}, \cite{Dabholkar:2010}. 
Due to the symmetric product structure, the Hilbert space
splits up into sectors with different number of twists. These twists can be thought of as an effective winding 
around the spacetime circle, and one identifies $L=2 \pi Q_{1}Q_{5} R$. 
The spacetime energy of a mode with quantum number $n$ is $E=n/R$. 

The entropy of this system at leading order follows from Cardy's formula $S=2 \pi \sqrt{Q_{1} Q_{5} n}$. The entropy 
formula is actually known to subleading order in each of the charges \cite{Castro:2008ys}, and this has been recently 
extended to an exact formula\footnote{More precisely, the formulas presented below are for a supersymmetric index. 
In \cite{Dabholkar:2010}, there is a careful discussion of the conditions under which the index and degeneracy are equal.  
Here, we shall make the standard assumption that due to the pairing of long multiplets, the index is the relevant quantity.}
\footnote{Similar formulas have appeared in the literature based on the computation of the entropy to first 
subleading order \cite{Cvitan:2007hu}.} 
in the limit when one of the charges is infinite and the other two are finite \cite{Dabholkar:2010}. 
This formula is derived using the known structure and symmetries of the $\CN=4$ partition function, and the 
relation between the $4d$ and $5d$ theories:

\vspace{0.2cm}

\ndt {\it Entropy formulas for the five dimensional theories with three charges $(Q_{1},Q_{5},n)$}
\bea \label{entropyexact5d}
S = 2 \pi \sqrt{Q_{1}Q_{5}n} \ , && \qquad n \to \infty , \quad  (Q_{1},Q_{5}) \quad  {\rm fixed} \ ; \\
S = 2 \pi \sqrt{Q_{1}Q_{5}(n+3)} \ , && \qquad Q_{1} \to \infty , \quad  (Q_{5} , n) \quad  {\rm fixed} \, . \label{entropyexact5d2}
\eea
The first of these is in a Cardy limit in the IIB description with $c^{\rm IIB} = Q_{1} Q_{5}$, $L^{\rm IIB}_{0} = n$, while the second can be 
thought of as a Cardy limit in the dual heterotic description with $c^{\rm Het} =  Q_{5}(n+3)$, $L^{\rm Het}_{0} = Q_{1}$. 
The extra symmetries of the $\CN=4$ theory thus give us some new knowledge of the low energy conformal field theory 
describing the heterotic system. 

The entropy formula in the four dimensional theory is also known, in this case there is an spacetime electric-magnetic 
duality relating the two different descriptions and we have:

\vspace{0.2cm}

\ndt {\it Entropy formulas for the four dimensional theories with four charges $(Q_{1},Q_{5},n,K)$}
\bea \label{entropyexact4d}
S = 2 \pi \sqrt{(Q_{1}Q_{5}+4)nK} \ , && \qquad n \to \infty , \quad  (K, Q_{1},Q_{5}) \quad  {\rm fixed} \ ; \\
S = 2 \pi \sqrt{Q_{1}Q_{5}(nK+4)} \ , && \qquad Q_{1} \to \infty , \quad  (K, Q_{5} , n) \quad  {\rm fixed} \, . \label{entropyexact4d2}
\eea
Here we discuss $K=1$. In all the $4d$ and $5d$ formulas, we can also introduce another charge corresponding 
to the angular momentum of the five-dimensional black hole, we will not do so here.

Now, in any theory with physical energy $E = n/R$ and a Cardy-like entropy-energy relation: 
\be\label{cardylike}
S = 2 \pi \sqrt{ E R c/6 + b} \ , 
\ee
we can derive a relation between the entropy and the temperature: 
\be\label{entemprel}
\frac{1}{T} = \frac{\pa S}{\pa E}  = \frac{2 \pi R c/6}{2 \sqrt{ E R c/6 + b} } 
\ee
or: 
\be\label{STrelgen}
S = \pi \,  (2 \pi c R/6) T \ . 
\ee
We can apply this to our theory of interest since the entropy formulas  \eqref{entropyexact5d}, 
\eqref{entropyexact4d} are indeed Cardy-like.  The relation \eqref{STrel}  holds for the three-charge 
and the four-charge system in {\it all} ranges of charges as long as one of the charges is very large. 
In this case, $c/6=Q_{1}Q_{5}$ which is the effective winding number of the long string, and so we get:
\be\label{STrel}
S = \pi \,  L \, T \ ,
\ee
consistent with the general relation \eqref{STrel1dgas} with $c_{\rm eff}=6$ as is the case for the D1-D5 system.


\subsection{The decay rate for large but finite charges} \label{microcorrections}

In this subsection, we will summarize the calculation of \cite{Das:1996wn}, keeping  however, the charges large but finite. 
We take the system to be away from extremality at a non-zero temperature $T_{H}$ as in the previous sections. 
We further make the dilute gas approximation\footnote{Here, we shall still keep $n \gg 1$ so that we can consider 
the BPS string as a large reservoir. We shall therefore only consider the first correction to the leading order formula. 
More generally, we can consider $n$ to be finite and use the heterotic description of the system. In that case, we shall 
be in the Cardy limit as mentioned above, so we can use the exact formulas \eqref{entropyexact5d2},\eqref{entropyexact4d2}.} 
$n \ll Q_{1}Q_{5}$ which implies that the typical amplitude of oscillations of the effective string is smaller 
than the typical wavelength of oscillations. This allows us to ignore the interactions between the left and 
right movers \cite{Das:1996wn, Maldacena:1997cg}. 

In this case, one has the decoupled left and right moving theories with their own effective thermal distributions 
governed by the temperatures $T_{L}$ and $T_{R}$ with: 
\be\label{leftrighttemp}
\frac{1}{T_{H}} = \half \left( \frac{1}{T_{L}} + \frac{1}{T_{R}} \right) \ .
\ee 
The temperatures $T_{L,R}$ are determined by demanding that the total number of left and right movers are fixed. 
The thermodynamic relations found in the previous subsection all hold for the chiral theories as well, in 
particular: 
\be\label{STrel1dgaschiral}
S_{L,R} = \pi L T_{L,R} \ . 
\ee

The amplitude for annihilation of two massless open string states into a graviton can be read off from the 
Born-Infeld action or from a disk diagram\footnote{One should really average over 
the coupling of the various operators in the SCFT. As was shown in \cite{Avery:2009tu}, in this limit, 
the average can be replaced by the coupling to a single operator. A more careful computation of the one-point 
function from first principles remains to be done. We thank S. Mathur for emphasizing this point.} \cite{Hashimoto:1996kf}. 
One chooses static gauge and identifies 
the two dimensions of the effective string to be $X^{0},X^{1}$ of spacetime. If the open strings carry momenta 
$(p_{0},p_{1})$, $(q_{0},q_{1})$ and the graviton has momentum $(k_{0},k_{1}, \vec{k})$ where $\vec{k}$ 
is the momentum in the directions transverse to the string, the amplitude is given by\footnote{The $\sqrt{2}$ is a 
symmetry factor for the off-diagonal graviton $h_{67}$.}: 
\be\label{amp}
\CA = \sqrt{2} \,  p \cdot q \ .
\ee
Putting in the kinematic factors, 
the decay rate for a pair of open strings to produce a closed string with $k_{1}=0$ is: 
\be\label{decayGamma}
\G(p,q,k) = \frac{\kappa_{d}^{2} \, 4 \pi^{2}}{4 L}  \delta(p_{0} + q_{0} - k_{0})  \delta(p_{1} + q_{1} - k_{1}) 
\frac{|\CA|^{2}}{p_{0} q_{0} k_{0}} \frac{d^{d}k}{(2 \pi )^{4}} \ , 
\ee
where $d=4,5$, $L=2 \pi Q_{1} Q_{5} R$. The constants 
$\kappa_{5}^{2} = \kappa^{2}/2 \pi R V_{T^{4}} $  and $\kappa_{4}^{2} = \kappa_{5}^{2}/2 \pi \wt R$ where $V_{T^{4}}$ is the volume of the compact $T^{4}$, and $R$ and $\wt R$ are the radii of the two circles $S^{1}$ and $\wt S^{1}$. 

To compute the total decay rate, one has to average over initial configurations:
\be\label{decayint}
\Gamma(k) = \int_{-\infty}^{\infty}  \frac{L dp_{1}}{2 \pi}  \int_{-\infty}^{\infty}  \frac{L dq_{1}}{2 \pi} \, \G(p,q,k) \, \rho(q_{0},q_{1}) \,  \rho(p_{0},p_{1}) \ , 
\ee
where $\rho(p_{0},p_{1})$ is the thermal distribution function with temperature $T_{H}$. 
We assume the graviton has no momentum along the string direction, so $p_{1} + q_{1} =0$, and since the open strings 
are massless, $p \cdot q = p_{0} q_{0} - p_{1} q_{1} = 2 |p_{1}|^{2}$.  Using the above expression \eqref{decayGamma}, one finds:
\be\label{totaldecay}
\G(k) = \frac{\kappa_{d}^{2} L }{ k_{0}} \left(\frac{k_{0}}{2} \right)^{2} \rho_{L}(k_{0}/2) \, \rho_{R}(k_{0}/2) \, \frac{d^{d}k }{(2 \pi)^{d}} \ , 
\ee
where $\rho_{L,R}$ are the thermal distribution functions of the left and right movers with temperatures $T_{L,R}$. Expressing 
the various quantities in terms of the charges and the energy of the graviton, we have:
\be\label{totaldecay2}
\G(k) = \kappa_{d}^{2}  L \,   \frac{\o}{4} \,  \rho_{L}(\o/2) \, \rho_{R}(\o/2) \, \frac{d^{d}k }{(2 \pi)^{d}} \ . 
\ee
Using the detailed balance condition:
\be\label{detbal}
\G (\o) = \s_{\rm abs} (\o) \frac{1}{e^{\o/T_{H}} -1} \, \frac{d^{d}k }{(2 \pi)^{d}} \ , 
\ee
we can read off the absorption cross section:
\be\label{abscross}
 \s_{\rm abs} (\o)  =  \kappa_{d}^{2}  L \,   \frac{\o}{4} \,  \frac{e^{\o/T_{H}} -1}{(e^{\o/2T_{L}} -1)(e^{\o/2T_{R}} -1)}  \ . 
\ee
We are interested in the limit $T_{H} \to 0$, which implies  
$T_{R} \ll T_{L}$ and  $T_{H} \approx 2 T_{R}$. 
In the low energy limit $\w \ll  T_{L}$,  $\rho_{L}(\o/2) \approx 2 T_{L}/\o$, so the cross 
section reduces to: 
\be\label{abscrosslim}
 \s_{\rm abs}   = \frac{\kappa_{d}^{2}}{2}  L \, T_{L} \ .
\ee
Plugging in the entropy-temperature relation \eqref{entemprel} for the left movers, and noting that $\k_{d}^{2} = 8 \pi G_{d}$ we get:
\be\label{absgen}
\s_{\rm abs}  = 4 G_N S_{L} \ . 
\ee
All these calculations apply equally for four and five dimensions. 

As a final note, we compare the regimes of validity of the microscopic and macroscopic calculations.
Both of them are performed with control when the charges obey\footnote{As we mentioned, one can push 
the microscopic one to the case where $n$ is finite.}:  
\be \label{scaling}
1 \ll n \ll Q_{1} Q_{5}  \, .
\ee
 In addition, the microscopic 
D-brane regime needs $g_{s}^{2}n, g_{s}^{2} Q_{1} Q_{5}  \ll 1$, while the gravity calculation 
requires the opposite one \eqref{sugraapprox}. To go between the two, we need to tune the modulus $g_{s}$.

\section{Discussion} \label{discuss}

Our calculations throw up interesting issues and new directions to pursue, we list some of them here. Some of these 
issues are currently being studied \cite{Kuperstein:2010pw}.
\begin{enumerate}

\item The agreement of the leading order scattering calculations in the  D1-D5 system was explained by 
the non-renormalization theorem of certain quantities in the underlying SCFT \cite{Maldacena:1996iz}. 
This argument\footnote{In this context, recent developments have been made in \cite{Dabholkar:2007ey}, \cite{deBoer:2008ss}.} 
needed $(4,4)$ supersymmetry on the SCFT, and did not really rely on the large charge limit.  
Our calculation for the five dimensional black hole 
can be thought of as an explicit computation of these quantities in the SCFT. For the four dimensional case, our 
results point to similar results in the 
underlying $(0,4)$ SCFT. It would be interesting to come up with  non-renormalization arguments in this case.

\item Our microscopic derivation was fairly general in that we only assumed an effective string description. With this assumption, 
we get an answer $\s_{\rm abs} = 4 G_N S$. 
This suggests that we should look for arguments in the macroscopic theory which relies on 
an underlying $AdS_{3}$. 
Relatedly, the $AdS_{3}$ helped in writing an exact entropy formula (to all orders in $\a'$). 
Can we do the same for $\s_{\rm abs}$? 

\item The entropy calculations actually agree in the two pictures even beyond any kind of Cardy limit, and there 
is an exact transcendental function which captures the full answer. It would be interesting to see if this 
can be done for the scattering cross section.

\item It would be very interesting to study the small black hole \cite{Dabholkar:2004yr} case using similar techniques as ours. 

\item It is of course important which field we use to scatter. Here we use the scalar which comes from dimensional 
reduction of the ten dimensional metric. What is the answer for other scalars? 
Other amplitudes like scattering of vector, graviton
 are also interesting to calculate. Even if they vanish to leading order, 
we can still compare the rate of vanishing and the coefficients \cite{Emparan:1997iv}. 

\item Finally, it is interesting to note that the form of our answer$\s_{\rm abs} = 4 G_N S_{\rm micro}$ 
looks a lot like the relation $\eta = s/4 \pi$ between 
the viscosity and the entropy density in the fluid-gravity correspondence \cite{Son:2007vk}. 
Since we are computing the two point function 
of an off-diagonal transverse graviton, it is indeed very similar on the gravity side of the calculation. Since we are in 
asymptotically flat space, it is not clear if there is a dual fluid description as in the AdS context -- note however the recent 
work \cite{Bredberg:2010ky}. It would be interesting to make the various connections precise. 
In particular, this ratio was found to change upon including higher derivative corrections in $AdS$ space. It would be interesting 
to see if our calculation can be generalized to other extremal black holes in asymptotically flat space. 

\end{enumerate}

\section*{Acknowledgements}
It is a pleasure to thank useful and enjoyable discussions with Atish Dabholkar, Gautam Mandal, 
Shiraz Minwalla, Boris Pioline, Mukund Rangamani and especially Ashoke Sen. 
We would like to thank Ofer Aharony, Sumit Das and Samir Mathur for critical comments on a preliminary  draft. 
S.~M. would like to thank IAFE and the University of Buenos Aires for local hospitality during a visit when part of this work 
was completed. 
The works of S.~M. and S.~K. are supported in part by the European Commission Marie Curie Fellowship under the
contracts PIIF-GA-2008-220899 and IEF-2008-237488 respectively.

\bibliographystyle{utphys}
\bibliography{BHScattering}

\end{document}